\documentclass[conference]{IEEEtran}
\IEEEoverridecommandlockouts
\usepackage{amsmath,amssymb,amsfonts}
\usepackage{graphicx}
\usepackage{bm}
\usepackage[export]{adjustbox}
\usepackage{hyperref}
\usepackage{dcolumn}
\usepackage{algorithm}
\usepackage{algpseudocode}
\usepackage{siunitx}
\usepackage{tikz}

\begin{document}

\title{Stochastic Gradient-Descent Calibration of Pyragas Delayed-Feedback Control for Chaos Suppression in the Sprott Circuit}
\author{
    \IEEEauthorblockN{1\textsuperscript{st} Adib Kabir}
    \IEEEauthorblockA{\textit{Department Of Physics} \\
    \textit{Gettysburg College} \\
    Gettysburg, PA 17325, USA \\
    kabiad01@gettysburg.edu}
    \and
    \IEEEauthorblockN{1\textsuperscript{st} Onil Morshed}
    \IEEEauthorblockA{\textit{Department Of Computer Science} \\
    \textit{Gettysburg College} \\
    Gettysburg, PA 17325, USA \\
    morson01@gettysburg.edu}

    \and 
    \IEEEauthorblockN{1\textsuperscript{st} Oishi Kabir}
    \IEEEauthorblockA{\textit{Department Of Mathematics} \\
    \textit{Gettysburg College} \\
    Gettysburg, PA 17325, USA \\
    kabioi01@gettysburg.edu}

    \and 
    \IEEEauthorblockN{1\textsuperscript{st} Juthi Hira}
    \IEEEauthorblockA{\textit{Department of Mechanical Engineering,} \\
    \textit{ Bangladesh University Of Engineering Technology (BUET)} \\
    Dhaka, Bangladesh \\
    2010004@me.buet.ac.bd}

    \and
    \IEEEauthorblockN{2\textsuperscript{nd} Dr. Caitlin Hult}
    \IEEEauthorblockA{\textit{Department Of Mathematics,} \\
    \textit{Faculty Of Mathematics, Gettysburg College} \\
    Gettysburg, PA 17325, USA \\
    chult@gettysburg.edu}
}

\maketitle

\begin{abstract}
This paper explores chaos control in the Sprott circuit by leveraging Stochastic Gradient Descent (SGD) to calibrate Pyragas delayed feedback control. Using a third-order nonlinear differential equation, we model the circuit and aim to suppress chaos by optimizing control parameters (gain $K$, delay $T_{\text{con}}$) and the variable resistor $R_v$. Experimental voltage data, extracted from published figures via WebPlotDigitizer, serve as the calibration target. We compare two calibration techniques: sum of squared errors (SSE) minimization via grid search and stochastic gradient descent (SGD) with finite differences. Joint optimization of $K$, $T_{\text{con}}$, and $R_v$ using SGD achieves superior alignment with experimental data, capturing both phase and amplitude with high fidelity. Compared to grid search, SGD excels in phase synchronization, though minor amplitude discrepancies persist due to model simplifications. Phase space analysis confirms the model ability to replicate the chaotic attractor geometry, despite slight deviations. We analyze the trade-off between calibration accuracy and computational cost, highlighting scalability challenges. Overall, SGD-based calibration demonstrates significant potential for precise control of chaotic systems, advancing mathematical modeling and applications in electrical engineering.
\end{abstract}


\section{Introduction}
Chaotic systems, characterized by extreme sensitivity to initial conditions and unpredictable long-term behavior, play a critical role in electrical engineering, with applications in secure communications \cite{Cuomo1993}, power electronics \cite{Banerjee1998}, neuromorphic computing \cite{Schuman2017}, and signal processing \cite{Pecora1990}. These systems, driven by nonlinear dynamics, present both opportunities and challenges: chaotic signals enhance encryption by generating complex, unpredictable patterns \cite{Cuomo1993}, yet their instability can disrupt systems requiring precise control, such as power grids \cite{Banerjee1998}. Consequently, developing robust methods to control chaos is essential for harnessing its benefits while ensuring system stability, with implications for fields ranging from cryptography to biomedical engineering.

The study of chaos control has a rich history, beginning with the seminal work of Ott, Grebogi, and Yorke (OGY), who introduced a method to stabilize unstable periodic orbits (UPOs) using small, targeted perturbations \cite{Ott1990}. Pyragas advanced this field in 1992 with delayed feedback control, a non-invasive technique that stabilizes UPOs by applying a feedback signal proportional to the difference between current and delayed states, vanishing upon stabilization \cite{Pyragas1992}. This approach has been widely adopted due to its simplicity and effectiveness \cite{Scholl2008}. Other methods, such as adaptive control \cite{Yang1995} and synchronization techniques \cite{Boccaletti2000}, have further expanded the toolkit for managing chaotic dynamics, addressing challenges like parameter sensitivity and real-time implementation.

The Sprott circuit, introduced by Sprott in 2000 \cite{Sprott2000}, is a minimal electronic system that generates chaotic dynamics using basic components like resistors, capacitors, diodes, and operational amplifiers. Its simplicity and accessibility make it an ideal testbed for chaos control studies. In recent year, Merat et al. \cite{Merat2007} applied Pyragas delayed feedback control to the Sprott circuit, modeling its dynamics with a third-order nonlinear differential equation and experimentally validating chaos suppression. However, their raw data was unavailable, requiring us to extract voltage traces from published figures using WebPlotDigitizer \cite{Rohatgi2021}. While their work established a foundation, it relied on traditional grid search methods for parameter calibration, which are computationally intensive and scale poorly with increasing parameter dimensionality \cite{Nocedal2006}.

Calibrating chaotic systems is inherently challenging due to their sensitivity to initial conditions and parameters, where small deviations can lead to significant trajectory divergence \cite{Strogatz2018}. Traditional optimization methods, such as grid search, are limited by their exhaustive nature, making them impractical for high-dimensional problems \cite{Nocedal2006}. In contrast, stochastic gradient descent (SGD), a cornerstone of artificial intelligence and machine learning, offers a scalable, iterative approach to parameter optimization by approximating gradients using mini-batches of data \cite{Goodfellow2016}. While SGD has been successfully applied to dynamical systems, such as neural network parameter estimation \cite{Kingma2015}, its use in calibrating chaotic electronic circuits remains largely unexplored. Recent studies have begun to bridge this gap, applying machine learning techniques to chaotic systems \cite{Pathak2018}, but few have focused on integrating SGD with delayed feedback control for circuits like the Sprott system.

This study introduces a novel AI-based approach by leveraging Stochastic Gradient Descent to calibrate Pyragas delayed feedback control for the Sprott circuit, building on the model by Merat et al. \cite{Merat2007} without reproducing their experimental setup. We hypothesize that SGD’s iterative, gradient-based optimization can outperform traditional grid search in achieving precise alignment between simulated and experimental dynamics, particularly in phase synchronization. Our objectives are threefold: (1) to simulate the circuit’s dynamics using the third-order differential equation, (2) to calibrate control parameters ($K$, $T_{\text{con}}$) and the variable resistor $R_v$ using SSE and SGD, and (3) to evaluate their performance against extracted experimental data. By demonstrating SGD’s superior performance in phase synchronization and analyzing the trade-off between accuracy and computational cost, this work highlights the transformative potential of AI-driven methods in modeling and controlling chaotic circuits. The findings contribute to nonlinear dynamics and electrical engineering, with applications in secure communications \cite{Cuomo1993}, power systems \cite{Banerjee1998}, and educational platforms \cite{Sprott2000}, paving the way for more efficient and scalable chaos control strategies. Code is available in the supplemental material section.

\section{Theory}
The Sprott circuit, a minimal electronic system designed to exhibit chaotic dynamics, leverages a simple configuration of resistors, capacitors, diodes, and operational amplifiers, as outlined by Merat et al. \cite{Merat2007}. The circuit’s nonlinearity, essential for chaos, arises from a diode-based feedback subcircuit \( D(x) \), implemented with a pair of diodes and an operational amplifier to approximate a piecewise-linear function. Building on this foundation, our study adopts the third-order nonlinear differential equation established by Merat et al. \cite{Merat2007}, expressed in dimensionless time \(\tau = \tilde{t} / (RC)\) (where \(\tilde{t}\) is physical time and \(RC\) is the characteristic time scale) as follows:
\begin{equation}
\dddot{x}(\tau) + \frac{R}{R_v} \ddot{x}(\tau) + \dot{x}(\tau) - D(x(\tau)) = -\frac{R}{R_0} \left( V_0 + u(\tau) \right), \label{eq:merat_ode}
\end{equation}
where \( x(\tau) \) represents the capacitor voltage, and \(\dot{x}(\tau)\), \(\ddot{x}(\tau)\), and \(\dddot{x}(\tau)\) denote the first, second, and third derivatives with respect to \(\tau\), respectively. The parameters, as defined by Merat et al. \cite{Merat2007}, include: \( R = 47\,\text{k}\Omega \) (feedback loop resistance), \( R_v = 80\,\text{k}\Omega \) (tunable resistor for chaos control), \( R_0 = 157\,\text{k}\Omega \) (source resistance), \( C = 1\,\mu\text{F} \) (capacitance), and \( V_0 = 0.25\,\text{V} \) (input voltage). The nonlinear term \( D(x) \) is given by:
\begin{equation}
D(x) = -\min\left( \frac{R_2}{R_1} x, 0 \right), \label{eq:nonlinear_d}
\end{equation}
with \( R_1 = 15\,\text{k}\Omega \) and \( R_2 = 90\,\text{k}\Omega \), scaling negative inputs by a factor of 6 to introduce the asymmetry necessary for chaos \cite{Merat2007}.

To suppress chaos, we incorporate Pyragas delayed feedback control \cite{Merat2007}, augmenting the system with a control signal \( u(\tau) \), defined as:
\begin{equation}
u(\tau) = \operatorname{sat}(\tilde{u}(\tau)) + u_0, \label{eq:control_signal}
\end{equation}
where \( u_0 = 0 \) \cite{Merat2007}, and the feedback term is:
\begin{equation}
\tilde{u}(\tau) = K \left( x_1(\tau - T_{\text{con}}) - x_1(\tau) \right), \label{eq:feedback_term}
\end{equation}
with \( K \) as the gain and \( T_{\text{con}} \) as the delay time. The saturation function \(\operatorname{sat}(\cdot)\) limits the control signal to \( [-0.25, 0.25]\,\text{V} \) \cite{Merat2007}:
\begin{equation}
\operatorname{sat}(\tilde{u}(\tau)) = \begin{cases}
\tilde{u}(\tau) & \text{if } |\tilde{u}(\tau)| \leq \tilde{u}_{\text{max}}, \\
\tilde{u}_{\text{max}} \cdot \operatorname{sign}(\tilde{u}(\tau)) & \text{if } |\tilde{u}(\tau)| > \tilde{u}_{\text{max}}.
\end{cases} \label{eq:saturation}
\end{equation}
For analysis, we adopt the state-space formulation from Merat et al. \cite{Merat2007}, defining state variables \( x_1 = x \), \( x_2 = -\dot{x} \), and \( x_3 = \ddot{x} \), leading to the system of first-order equations:
\begin{equation}
\begin{aligned}
\dot{x}_1 &= -x_2, \\
\dot{x}_2 &= -x_3, \\
\dot{x}_3 &= -\frac{R}{R_v} x_3 + x_2 + D(x_1) - \frac{R}{R_0} \left( V_0 + u(\tau) \right). \label{eq:state_space}
\end{aligned}
\end{equation}
This representation enables the study of the circuit’s three-dimensional phase space dynamics, where \( D(x_1) \) and \( u(\tau) \) drive the chaotic trajectory. Our calibration effort focuses on optimizing \( K \), \( T_{\text{con}} \), and \( R_v \) to align simulated outputs with experimental data, building on the framework established by Merat et al. \cite{Merat2007} and Pyragas \cite{Pyragas1992}.

\section{Methodology}
\subsection{Data Extraction and Preprocessing}
A significant challenge in calibrating the Sprott circuit model was the absence of raw experimental data from Merat et al. \cite{Merat2007}. As no digital datasets were provided and attempts to contact the authors were unsuccessful, we manually extracted voltage trace data from published figures in the paper. Using \textit{WebPlotDigitizer}, a widely used tool for digitizing graphical data, we carefully plotted the signal \( x_1(t) \), representing the capacitor voltage in the Sprott circuit, from specific experimental figures, namely the phase diagram (Fig. 3f) and time series plots (Fig. 3d) within the paper \cite{Merat2007}. This manual process involved visually identifying and marking data points on each figure. This cumbersome data plotting task further introduced systematic errors in the experimental data. Despite these limitations, we obtained a dataset of 284 time-aligned points, which was saved in a CSV file and served as the experimental reference for our calibration step.

\begin{figure}[t]
\begin{minipage}[t]{1\columnwidth}
\centering
\begin{adjustbox}{width=\linewidth}
\includegraphics{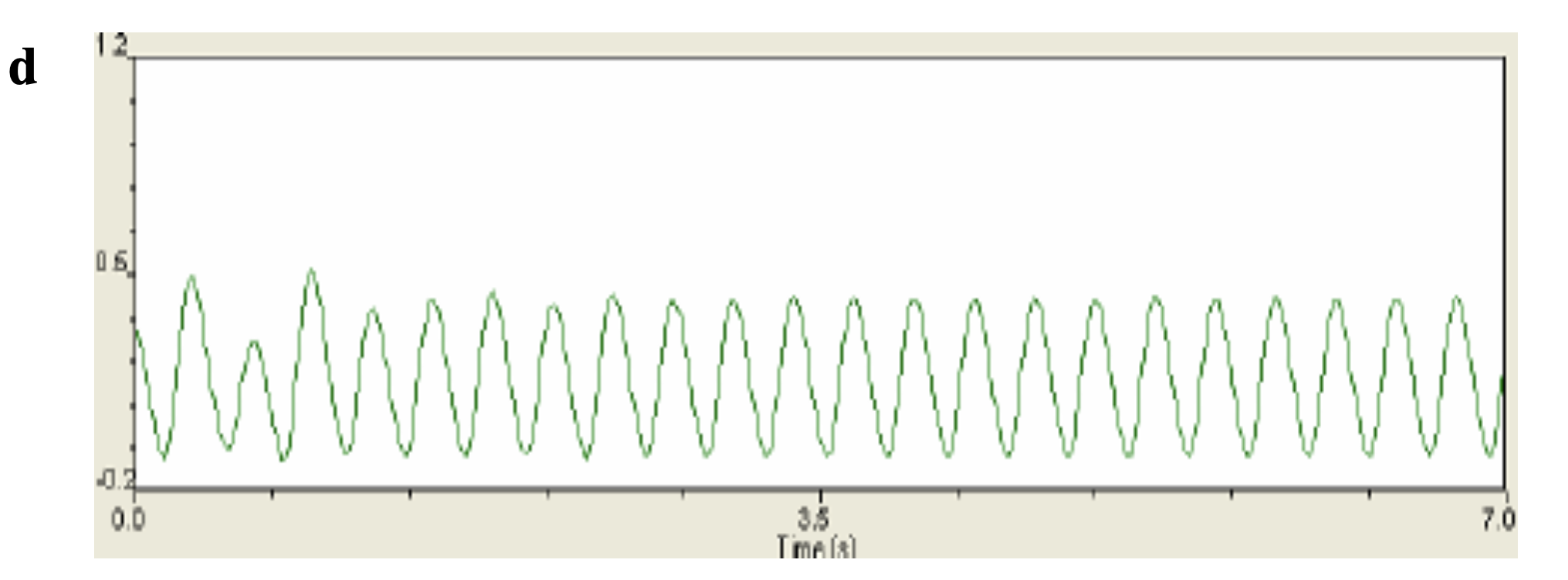}
\end{adjustbox}
\caption*{Experimental $x_1(t)$ from Merat et al. \cite{Merat2007}}
\end{minipage}\hfill
\\
\begin{minipage}[t]{1\columnwidth}
\centering
\begin{adjustbox}{width=\linewidth}
\includegraphics{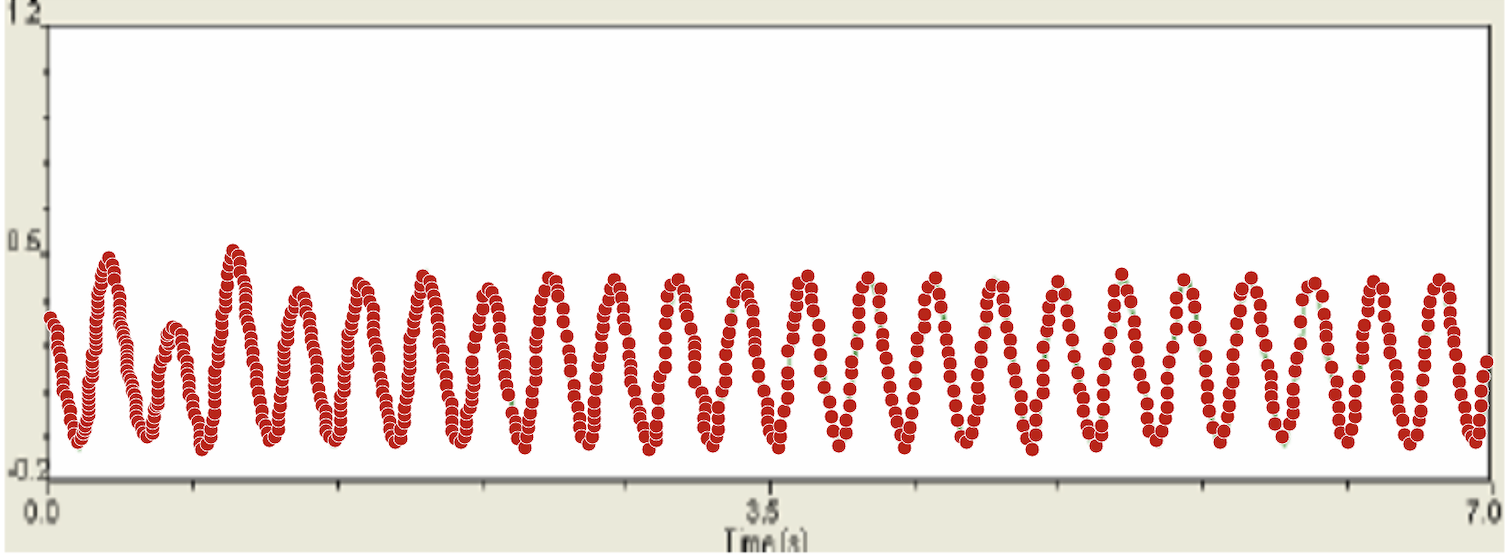}
\end{adjustbox}
\caption*{Digitized time series using WebPlotDigitizer}
\end{minipage}
\caption{Experimental data for the capacitor voltage \( x_1(t) \) extracted manually from published figures in \cite{Merat2007} using WebPlotDigitizer. (a) Original time series plot. (b) Digitized dataset of 284 time-aligned points.}
\label{fig:data_trace}
\end{figure}

\subsection{Calibration Using Sum of Squared Errors}
To calibrate the Pyragas delayed feedback control parameters for the Sprott circuit, we employed a sum of squared errors (SSE) traditional minimization approach. This method optimized the control parameters—time delay \( T_{\text{con}} \in \mathbb{R}^+ \), corresponding to the period of the target unstable periodic orbit (UPO), and gain \( K \in \mathbb{R} \)—to align the simulated capacitor voltage \( x_1(\tau) \) with the experimental data. The SSE loss function is defined as:
\begin{equation}
\mathcal{L}_{\text{SSE}}(T_{\text{con}}, K) = \sum_{i=1}^N \left( x_1^{\text{sim}}(\tau_i, T_{\text{con}}, K) - x_1^{\text{exp}}(\tau_i) \right)^2,
\end{equation}
where \( x_1^{\text{sim}}(\tau_i, T_{\text{con}}, K) \) is the simulated voltage at dimensionless time \( \tau_i \) and is obtained by numerically integrating the state-space equations from Section II. To numerically integrate, we used a fourth-order Runge-Kutta method with a fixed time step of \( \Delta \tau = 0.01 \). The experimental voltage \( x_1^{\text{exp}}(\tau_i) \) consists of \( 284 \) data points as evident from Fig.~\ref{fig:data_trace}.

We minimized \( \mathcal{L}_{\text{SSE}} \) via a grid search over the parameter space \( T_{\text{con}} \in [1.5, 2.5] \) and \( K \in [0.8, 1.5] \). We have chosen these ranges to encompass values K. Merat and their research group \cite{Merat2007} used in their simulations. The grid was discretized with a resolution of 100 points per dimension, yielding 10,000 evaluations of \( \mathcal{L}_{\text{SSE}} \). This was implemented in Python using \texttt{NumPy} for array operations and \texttt{SciPy} for numerical integration. The algorithm of our implementation is shown in Fig.~\ref{fig:grid_algo}.

\begin{figure}[t]
\caption{Grid Search Calibration Using Sum of Squared Errors (SSE).}
\label{fig:grid_algo}
\begin{algorithmic}[1]
\State \textbf{Input:} Experimental data \( \{x_1^{\text{exp}}(\tau_i)\}_{i=1}^{N} \), delay range \( T_{\text{con}} \in [1.5, 2.5] \), gain range \( K \in [0.8, 1.5] \), resolution \( R = 100 \)
\State \textbf{Initialize:} Define grid points
\[
T_{\text{grid}} = \{T_1, T_2, \dots, T_R\}, \quad K_{\text{grid}} = \{K_1, K_2, \dots, K_R\}
\]
\State Set best loss \( \mathcal{L}_{\min} \gets \infty \), best parameters \( (T^*, K^*) \gets (0, 0) \)
\For{each \( T \in T_{\text{grid}} \)}
    \For{each \( K \in K_{\text{grid}} \)}
        \State Simulate \( x_1^{\text{sim}}(\tau_i, T, K) \) using RK4 method with \( \Delta \tau = 0.01 \)
        \State Compute loss:
        \[
        \mathcal{L}_{\text{SSE}}(T, K) = \sum_{i=1}^N \left( x_1^{\text{sim}}(\tau_i, T, K) - x_1^{\text{exp}}(\tau_i) \right)^2
        \]
        \If{ \( \mathcal{L}_{\text{SSE}}(T, K) < \mathcal{L}_{\min} \) }
            \State \( \mathcal{L}_{\min} \gets \mathcal{L}_{\text{SSE}}(T, K) \)
            \State \( (T^*, K^*) \gets (T, K) \)
        \EndIf
    \EndFor
\EndFor
\State \textbf{Output:} Best parameters \( T^*, K^* \), with minimal loss \( \mathcal{L}_{\min} \)
\end{algorithmic}
\end{figure}

\subsection{Calibration with Stochastic Gradient Descent (SGD)}
In the context of mathematical optimization and artificial intelligence, Stochastic Gradient Descent (SGD) is a first-order iterative optimization algorithm used to minimize an objective function \( \mathcal{J}(\boldsymbol{\theta}) \), which typically represents a measure of error or loss. This method is widely used in optimization problems in machine learning and is particularly well-suited for chaotic dynamical systems where analytic gradients are difficult or impossible to compute. Let \( \mathcal{J} : \mathbb{R}^d \rightarrow \mathbb{R} \) be a real-valued objective Loss function defined over a \( d \)-dimensional parameter space. The goal of Stochastic Gradient Descent (SGD) aims to find out the minimum value of $\mathcal{J}(\boldsymbol{\theta}),$ where
\[
\boldsymbol{\theta} =
\begin{bmatrix}
\theta_1 \\
\theta_2 \\
\vdots \\
\theta_d
\end{bmatrix}
\in \mathbb{R}^d
\]
denotes the parameter vector. SGD is an iterative optimization algorithm defined by the following recursive definition:
\begin{equation}
\boldsymbol{\theta}^{(k+1)} = \boldsymbol{\theta}^{(k)} - \alpha_k \nabla_{\boldsymbol{\theta}}{\mathcal{J}}(\boldsymbol{\theta}^{(k)}),
\end{equation}
where:
\begin{itemize}
    \item \( \alpha_k > 0 \) is the learning rate parameter at iteration \( k \),
    \item \( \nabla_{\boldsymbol{\theta}} {\mathcal{J}}(\boldsymbol{\theta}^{(k)}) \) is an approximate gradient of the loss function, computed using only a small random sample, also known as mini-batch, of the full dataset.
    \item \( \boldsymbol{\theta}^{(k)} \) is the current guess for the parameter values at iteration \( k \).
\end{itemize}

To improve the fit between the simulation and the manually plotted experimental data of the Sprott circuit, we have implemented this calibration technique. For our model, we define the loss function \( \mathcal{J}(T, K, R_v) \) as follows:
\begin{eqnarray}
\mathcal{J}(T, K, R_v) &=& \sum_{i=1}^{N} \left( x_1^{\text{sim}}(t_i) - x_1^{\text{exp}}(t_i) \right)^2 \notag\\
&&+ \sum_{i=1}^{N} \left( u^{\text{sim}}(t_i) - u^{\text{exp}}(t_i) \right)^2
\end{eqnarray}
where \( x_1^{\text{sim}} \) and \( u^{\text{sim}} \) are outputs from the simulation, and \( x_1^{\text{exp}}, u^{\text{exp}} \) are the experimental values. Due to the complexity of gradient computation, we approximate gradients numerically using finite differences. For a small perturbation \( \varepsilon \), we estimate the partial derivatives:
\begin{eqnarray*}
\frac{\partial \mathcal{J}}{\partial T} &\approx& \frac{\mathcal{J}(T + \varepsilon, K, R_v) - \mathcal{J}(T, K, R_v)}{\varepsilon}, \\
\frac{\partial \mathcal{J}}{\partial K} &\approx& \frac{\mathcal{J}(T, K + \varepsilon, R_v) - \mathcal{J}(T, K, R_v)}{\varepsilon}, \\
\frac{\partial \mathcal{J}}{\partial R_v} &\approx& \frac{\mathcal{J}(T, K, R_v + \varepsilon_r) - \mathcal{J}(T, K, R_v)}{\varepsilon_r}
\end{eqnarray*}
with \( \varepsilon = 10^{-3} \), and \( \varepsilon_r = 10 \times \varepsilon \cdot 10^3 \) to reflect the scale of \( R_v \). For each parameter follows the standard SGD form, the update rule is given as follows:
\begin{equation}
\theta \leftarrow \theta - \alpha \frac{\partial \mathcal{J}}{\partial \theta}
\end{equation}
where \( \alpha = 0.01 \) is the learning rate and \( \theta \in \{T, K, R_v\} \). The algorithm of this calibration technique is mentioned below:

\begin{figure}[t]
\caption{SGD-Based Parameter Calibration.}
\label{fig:sgd_algo}
\begin{algorithmic}[1]
\State \textbf{Input:} Initial values for \( T, K, R_v \); learning rate \( \alpha \); number of iterations \( N \)
\State \textbf{Initialize:} Choose \( T, K, R_v \) within valid physical ranges
\For{each iteration \( n = 1 \) to \( N \)}
    \State Simulate the system using the current values of \( T, K, R_v \)
    \State Compute the loss \( \mathcal{J}(T, K, R_v) \)
    \State Estimate gradients using finite differences:
    \[
    \frac{\partial \mathcal{J}}{\partial T},\quad
    \frac{\partial \mathcal{J}}{\partial K},\quad
    \frac{\partial \mathcal{J}}{\partial R_v}
    \]
    \State Update each parameter using:
    \[
    \theta \leftarrow \theta - \alpha \frac{\partial \mathcal{J}}{\partial \theta}, \quad \text{for } \theta \in \{T, K, R_v\}
    \]
    \State \textbf{Enforce parameter limits:}
    \begin{itemize}
        \item If \( T < 6.0 \), set \( T = 6.0 \); if \( T > 7.0 \), set \( T = 7.0 \)
        \item If \( K < 0.05 \), set \( K = 0.05 \); if \( K > 0.25 \), set \( K = 0.25 \)
        \item If \( R_v < 60 \,\text{k}\Omega \), set \( R_v = 60 \,\text{k}\Omega \); if \( R_v > 100 \,\text{k}\Omega \), set \( R_v = 100 \,\text{k}\Omega \)
    \end{itemize}
\EndFor
\State \textbf{Output:} Calibrated parameters \( T, K, R_v \)
\end{algorithmic}
\end{figure}

\section{Results and Discussion}
To evaluate the effectiveness of our chaos control and calibration strategies for the Sprott circuit, we systematically analyzed the system’s behavior under various conditions: uncontrolled dynamics, phase space sweeps, and calibrated models using both sum of squared errors (SSE) and stochastic gradient descent (SGD). The results demonstrate the strengths and limitations of each approach in aligning simulated outputs with experimental data extracted from Merat et al. \cite{Merat2007}.

\subsection{A. Time Series Plots for the Uncontrolled Circuit}
To investigate the behavior of the Sprott circuit, we first simulated the system with \( R_v = 80 \, \text{k}\Omega \), aligning with the value employed in the experimental setup \cite{Merat2007}. The resulting time series for the three state variables—\( x_1(t) \), \( x_2(t) \), and \( x_3(t) \)—are detailed below. Figure 7 presents the time series plots of the Sprott circuit in its uncontrolled state, where no feedback control signal is applied, i.e., \( u(t) = 0 \). Under this condition, the governing third-order nonlinear differential equation simplifies to:
\[
\dddot{x}(t) + \frac{R}{R_v} \ddot{x}(t) + \dot{x}(t) - D(x(t)) = -\frac{R V_0}{R_0}.
\]

\begin{figure}[]
\centering
\begin{adjustbox}{width=\columnwidth}
\includegraphics{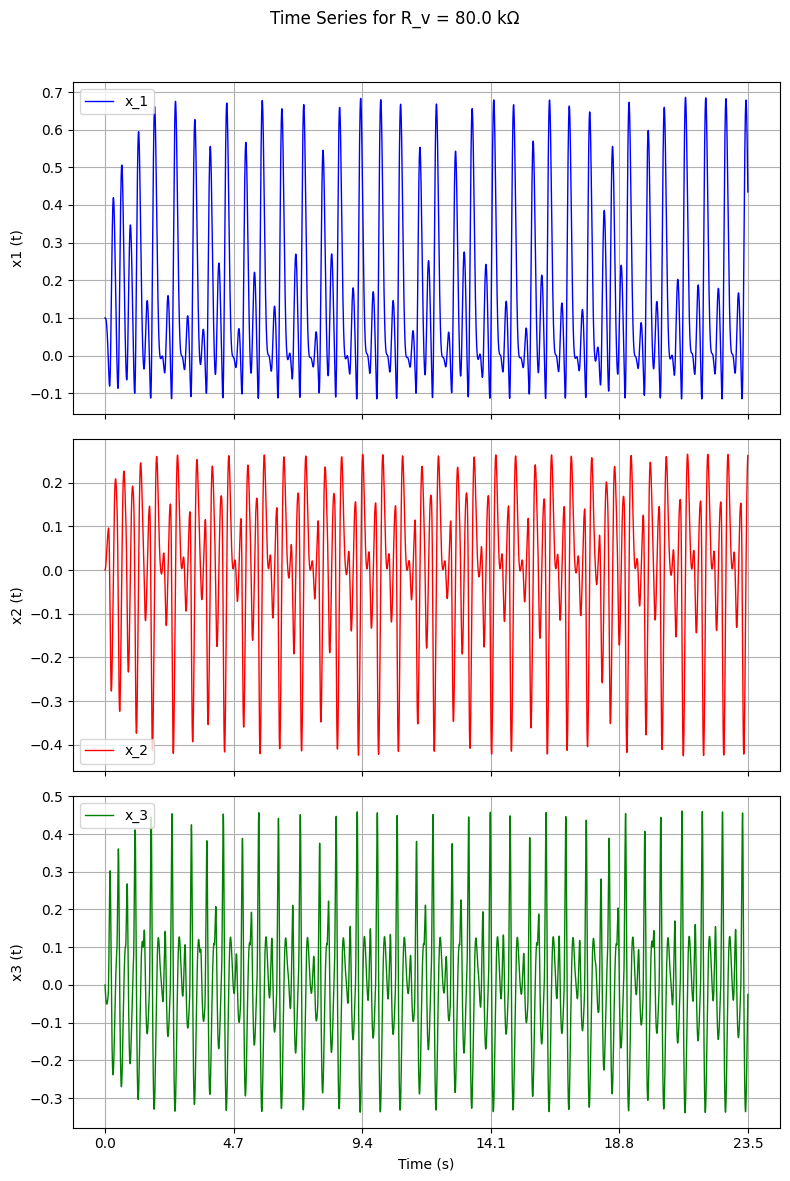}
\end{adjustbox}
\caption{Simulated time series for the Sprott circuit at \( R_v = 80.0\,\text{k}\Omega \).}
\label{fig:timeseries_80k}
\end{figure}
With \( u(t) = 0 \), the system evolves according to its intrinsic dynamics, free from external stabilization. The top plot in figure ~\ref{fig:timeseries_80k} illustrates \( x_1(t) \), representing the voltage across the capacitor. This signal reveals irregular yet bounded oscillations devoid of apparent periodicity, a hallmark of chaotic behavior. Its sensitivity to initial conditions and rapid temporal variations underscore the quintessential traits of nonlinear chaotic systems. In figure ~\ref{fig:timeseries_80k}, the middle plot depicts \( x_2(t) \), the negative first derivative of \( x_1(t) \), which corresponds to the circuit’s current. This trace exhibits sharp transitions and swift amplitude shifts, driven by the nonlinear diode element’s switching behavior. Meanwhile, the bottom plot in figure ~\ref{fig:timeseries_80k} portrays \( x_3(t) \), the negative second derivative of \( x_1(t) \), capturing the circuit’s swiftest responses. High-frequency oscillations and abrupt slope changes reflect the profound influence of the diode-induced nonlinearity, shaping the chaotic attractor’s structure.

\subsection{B. Phase Space Sweep for the Uncontrolled Circuit}
\begin{figure*}[t]
\centering
\begin{adjustbox}{width=\textwidth}
\includegraphics{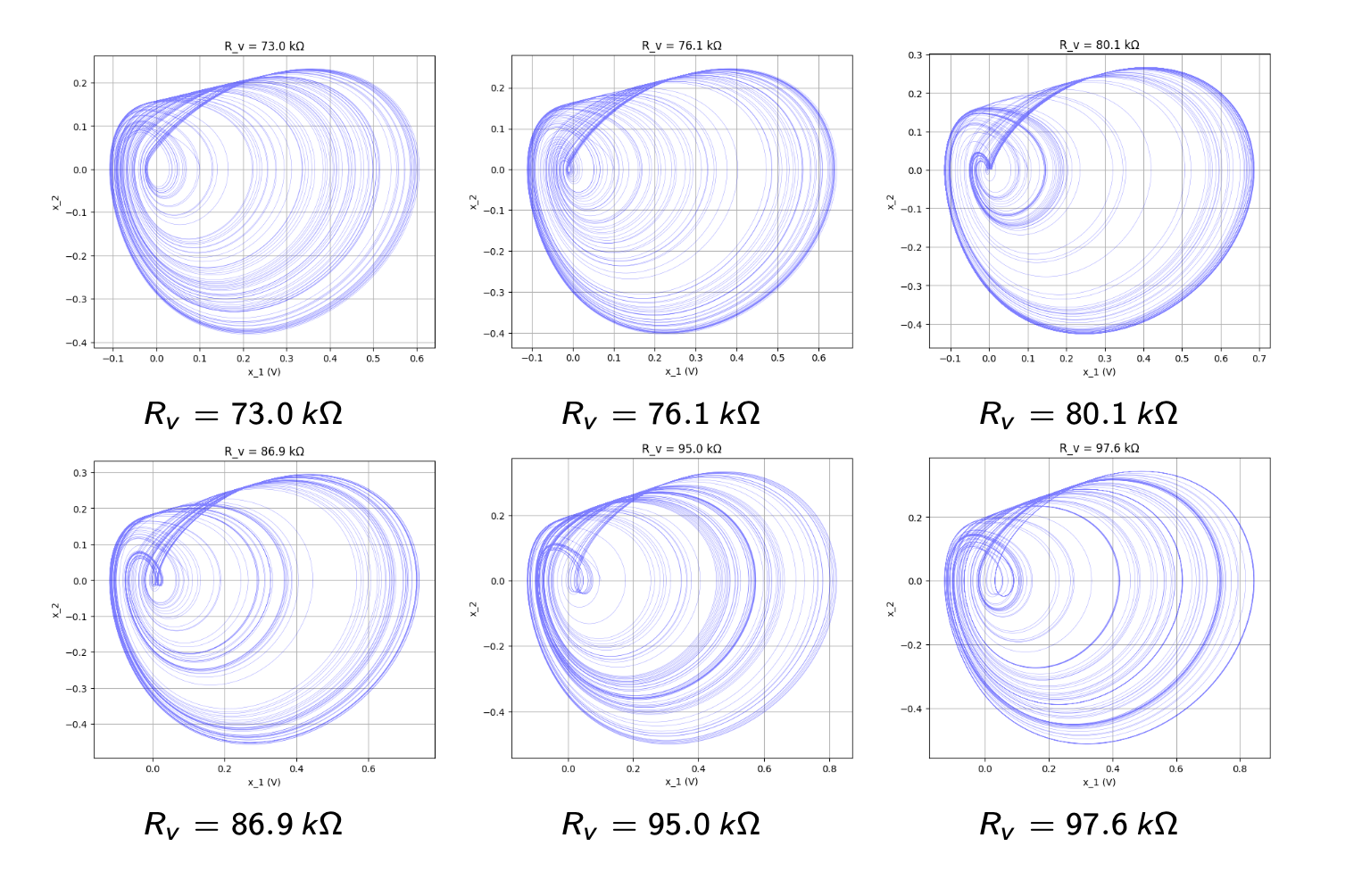}
\end{adjustbox}
\caption{Phase space plots for various values of \( R_v \) in the uncontrolled circuit. Each orbit reflects the system's evolution under a different resistance value, showing the sensitivity of the chaotic attractor to small parameter changes.}
\label{fig:phase_sweep}
\end{figure*}

In the absence of any control input—where \( u(t) = 0 \)—we explored the inherent behavior of the Sprott circuit by varying the variable resistor \( R_v \) across a range of values. This investigation involved simulating the circuit’s dynamics and generating phase portraits in the \( x_2 \) versus \( x_1 \) plane. As depicted in figure~\ref{fig:phase_sweep}, each subplot corresponds to a distinct \( R_v \) value, spanning from 73.0 k\(\Omega\) to 97.6 k\(\Omega\). These values were selected to probe the circuit’s behavior near the experimentally observed chaotic regime. For each \( R_v \) setting, the circuit equations were numerically integrated using the fourth-order Runge-Kutta method. The resulting trajectories in the \( x_1 \)-\( x_2 \) phase space display notable diversity. At lower \( R_v \) values, the attractors appear compact and tightly coiled, while higher values lead to stretched, increasingly complex orbits that eventually bifurcate, signaling transitions between distinct chaotic regimes. This phase space analysis underscores the system’s exquisite sensitivity to the resistor parameter \( R_v \). These plots offer critical insights for calibration and control strategies, revealing that even minor adjustments in \( R_v \) can profoundly alter the attractor’s geometry. Thus, precise parameter tuning emerges as essential for stabilizing or synchronizing chaotic behavior through feedback control methods. Together, these time series and phase sweep plots provide compelling evidence of the uncontrolled Sprott circuit’s chaotic nature, marked by the absence of regular patterns and a pronounced dependence on initial conditions. These traces establish a baseline for comparing against the calibrated models presented in subsequent sections, where feedback control and parameter optimization via Sum of Squared Errors (SSE) and Stochastic Gradient Descent (SGD) are implemented.

\subsection{Uncontrolled Simulation vs. Experimental Data}
\begin{figure}[t]
\centering
\begin{adjustbox}{width=\columnwidth}
\includegraphics{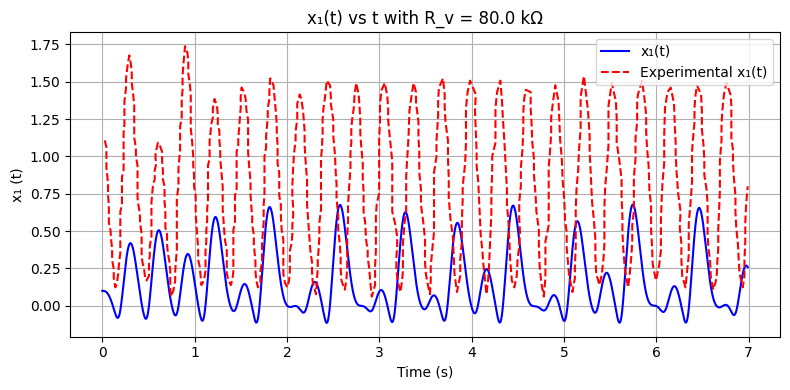}
\end{adjustbox}
\caption{Comparison of simulated (blue) and experimental (red dashed) capacitor voltage \( x_1(t) \) for \( R_v = 80.0\,\text{k}\Omega \) with no feedback control.}
\label{fig:sim_vs_exp_uncontrolled}
\end{figure}

To assess the fidelity of our uncontrolled Sprott circuit model, we compared the simulated capacitor voltage \( x_1(t) \) with the experimental data acquired via WebPlotDigitizer. The simulation was conducted by numerically integrating the third-order nonlinear system without feedback control, where \( u(t) = 0 \). As illustrated in figure~\ref{fig:sim_vs_exp_uncontrolled}, the simulated and experimental signals share several defining characteristics. Both exhibit aperiodic, amplitude-bounded behavior, affirming the presence of chaos and validating the model’s ability to capture the system’s general dynamics. However, notable discrepancies emerge: the simulated waveform displays lower peak amplitudes compared to the experimental data, and a gradual rightward phase drift is evident, indicating a temporal misalignment between the signals. Such mismatches are anticipated in chaotic systems, where even minute variations in parameters or initial conditions can precipitate divergent trajectories over time. This divergence is a natural consequence of modeling chaotic systems without control, suggesting that while the model’s structure is sound, achieving precise waveform replication necessitates the introduction of a control strategy. To address this, we incorporate a delayed feedback control signal \( u(t) \) to stabilize an unstable periodic orbit within the chaotic attractor, as outlined in Section I.

\subsection{SSE-Based Calibration}
\begin{figure}[t]
\centering
\begin{adjustbox}{width=\columnwidth}
\includegraphics{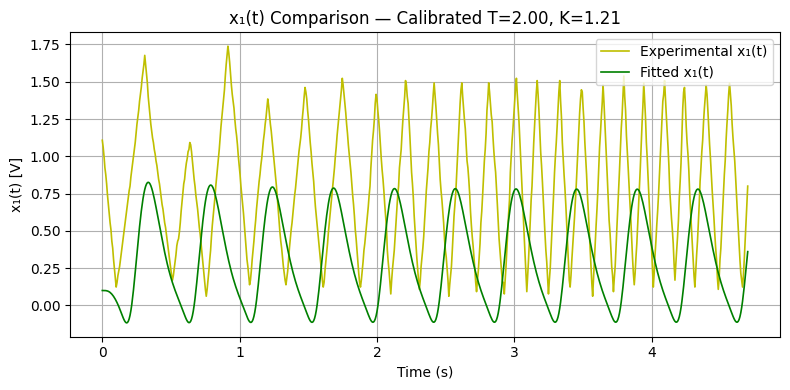}
\end{adjustbox}
\caption{Comparison of experimental \( x_1(t) \) (yellow) and SSE-calibrated simulation (green) with parameters \( T_{\text{con}} = 2.00 \), \( K = 1.21 \).}
\label{fig:sse_fit}
\end{figure}

To enhance the alignment between the simulated and experimental signals of the Sprott circuit, we employed the Sum of Squared Errors (SSE) calibration method introduced in Section III.A. This approach entails a grid search over predefined ranges of control parameters—specifically the delay \( T_{\text{con}} \) and the gain \( K \)—to minimize the pointwise discrepancy between the simulated voltage \( x_1(t) \) and its experimental counterpart. As depicted in figure~\ref{fig:sse_fit} , the SSE-calibrated simulation more accurately captures the general amplitude profile of the experimental waveform compared to the uncontrolled model. The overall envelope of oscillations aligns more closely, suggesting that the selected parameters effectively replicate the magnitude of the experimental voltage over time. Yet, a significant limitation persists: the simulation fails to synchronize with the experimental signal in terms of timing. The phase drift endures throughout the observed window, with simulated peaks occurring out of phase with their experimental counterparts. This arises because SSE focuses solely on minimizing local, pointwise errors, neglecting the global structure and inherent sensitivity of chaotic systems. Given that even slight differences in initial conditions or parameters can lead to rapid trajectory divergence in chaotic dynamics, this localized optimization does not ensure alignment of the overall trajectory—particularly its timing and long-term behavior—with the experimental data. This limitation underscores the need for more advanced calibration techniques, such as gradient-based or control-aware optimization methods, which we explore in the following section. \cite{Nocedal2006}.

\subsection{SGD Calibration with Fixed $R_v$}
\begin{figure}[t]
\centering
\begin{adjustbox}{width=\columnwidth}
\includegraphics{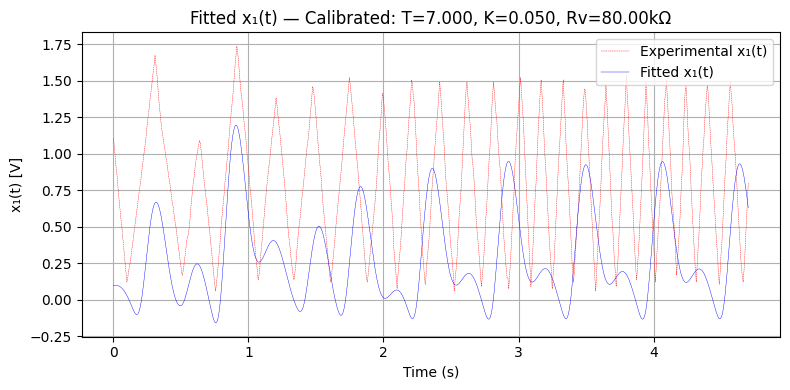}
\end{adjustbox}
\caption{Comparison of experimental \( x_1 \) (red dotted) and simulated (SGD-calibrated simulation (blue)) with fixed \( R_v = 80.0 \,\text{k}\Omega \), \( T_{\text{con}} = 7.000 \), \( K = 0.050 \).}
\label{fig:sgd_fixed}
\end{figure}

\begin{figure}[t]
\centering
\begin{adjustbox}{width=\columnwidth}
\includegraphics{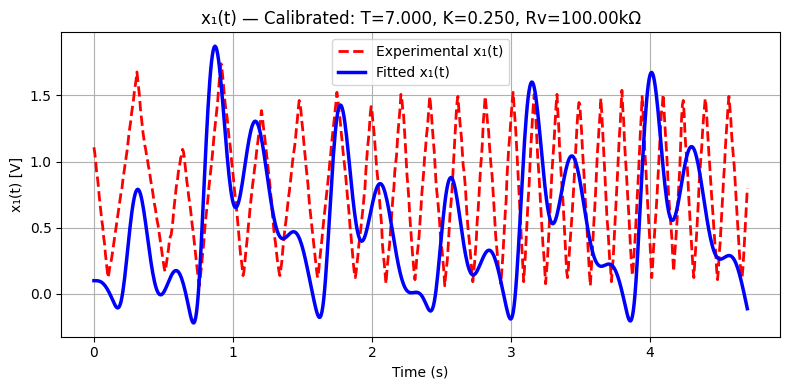}
\end{adjustbox}
\caption{Comparison of experimental \( x_1(t) \) (red dashed) and SGD-calibrated simulation (blue) with optimized \( T_{\text{con}} = 7.000 \), \( K = 0.250 \), and \( R_v = 100.0\,\text{k}\Omega \).}
\label{fig:sgd_variable}
\end{figure}

To address the shortcomings of the SSE method, we implemented stochastic gradient descent (SGD) with finite-difference gradient estimation. In this approach, we optimized only the feedback control parameters \( T_{\text{con}} \) and \( K \), maintaining the circuit parameter \( R_v \) at a fixed value of 80 k\(\Omega\), consistent with the experimental configuration by Merat et al. \cite{Merat2007}. As illustrated in figure~\ref{fig:sgd_fixed}, the fitted waveform demonstrates improved phase alignment and smoother oscillatory behavior in the early time window compared to the SSE-calibrated model. However, the amplitude remains underestimated, and synchronization deteriorates over time. These findings suggest that while adaptive tuning of control parameters enhances local dynamics, it falls short of overcoming the structural constraints imposed by a fixed \( R_v \). The circuit’s internal nonlinearities remain inadequately matched to the experimental system. Figure~\ref{fig:sgd_variable} shows that allowing \( R_v \) to vary yields the best fit among all models. The simulated signal now aligns well with the experimental waveform in both amplitude and phase, capturing peak sharpness and frequency content more accurately than in the fixed-\( R_v \) case by Fig.~\ref{fig:sgd_fixed}. The reduction in phase drift and improvement in waveform fidelity demonstrate that tuning internal system parameters like \( R_v \) is crucial for capturing the geometry of the underlying chaotic attractor phase. Nonetheless, this result highlights a fundamental insight: optimizing control parameters alone is insufficient in chaotic systems, where small structural mismatches can lead to divergence. Joint calibration of both control and physical parameters significantly improves synchronization. Yet some residual discrepancies persist, particularly in the amplitude of some high-frequency peaks. This suggests that further improvement may require expanding the parameter space of the loss function itself—for example, by introducing additional physical or nonlinear circuit parameters in the loss function—to better capture the order’s fine structure and enhance global trajectory matching.

\subsection{Phase Space Validation of Calibrated Parameters}
To validate the accuracy of the calibrated model, we analyzed the phase space behavior of the phase space by best-fit simulation obtained via simulation, using best-fit parameters \( T_{\text{con}} = 7.000 \), \( K = 0.250 \), and \( R_v = 100.0\,\text{k}\Omega \). This complements the time-domain comparison by evaluating whether the simulated trajectory correctly reproduces the geometry of the experimental attractor in the \( x_2 \)–\( x_1 \) plane.

\begin{figure}[t]
\centering
\begin{adjustbox}{width=\columnwidth}
\includegraphics{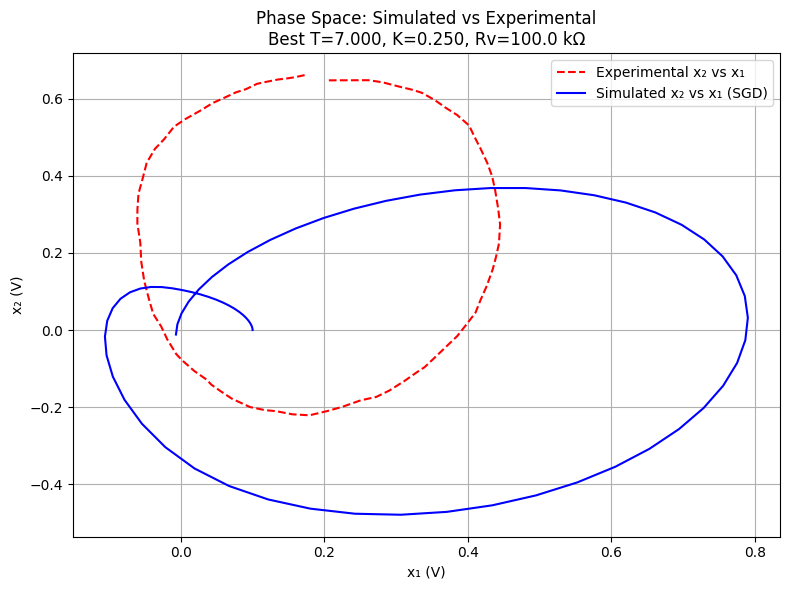}
\end{adjustbox}
\caption{Phase space comparison between experimental (red dashed) and simulated (blue) trajectories in the \( x_2 \)–\( x_1 \) plane using the best-fit parameters from SGD.}
\label{fig:phase_validation}
\end{figure}

As shown in Fig.~\ref{fig:phase_validation}, the simulated phase portrait captures the overall dynamics of the spatial phase space: the general loop shape, curvature, and bounding region are well aligned. This phase-space alignment indicates suggests that our model successfully faithfully reproduces the dynamics of the chaotic motion. However, subtle differences are noticeable. The simulated orbit is wider and less compact than the experimental trace. These discrepancies may arise from slight mismatches in fine-scale nonlinearities, diode characteristics, behavior, or unmodeled experimental constraints not fully captured by our simplified model. Another potential source of error could be that our loss function was optimized for time-series alignment and may not directly penalize differences in trajectory curvature or loop geometry. Expanding the loss function to include additional terms—such as derivatives or phase space metrics—could further reduce these discrepancies by better aligning the order structure.

In summary, this phase space analysis provides strong confirmation of the order’s validity, reinforcing the earlier time-domain findings while also pointing to areas for potential improvement. At the same time, it reveals that fully capturing the chaotic structure may require a more complex model or an enhanced cost function that accounts for geometric errors in state space.

\subsection{Remarks on Parameter Dimensionality and Experimental Cost}
Calibrating the circuit parameter \( R_v \) in addition to the feedback parameters \( T_{\text{con}} \) and \( K \) significantly improved the alignment between simulation and experimental data, as seen in Fig.~\ref{fig:sgd_variable}. This enhancement results from expanding the simulation parameter space from two to three dimensions, enabling the optimization algorithm’s algorithm to better capture the dynamics of the system. However, this comes at a significant computational cost.

As detailed in Appendix A, our finite-difference implementation approach requires two numerical simulations per parameter per simulation, for each gradient estimate. Increasing order the number of parameters from \( d = 2 \) to \( d = 3 \) raises the simulation count from \( 2 \times 2 = 4 \) to \( 2 \times 3 = 6 \) per simulation, a 67\% increase in simulations per epoch. Across 100 epochs, simulations, the total simulation count grows from 400 to 6000, increasing time complexity proportionally.

If additional loss terms were introduced in the cost function, the parameter count \( d \) would rise to \( d + k \) simulations, and the total simulation count would scale linearly with \( d \). While this richer parameter space enables a more accurate faithful reproduction of the chaotic order, it also introduces higher computational complexity, costs, and potential convergence issues. Thus, any expansion of the loss function must be carefully justified by expected gains in model accuracy and synchronization.

\section{Limitations}
Despite the model’s success in modeling experimental order under controlled conditions, several limitations exist. First, the manual data extraction via WebPlotDigitizer introduces uncertainty in resolution and alignment due to unavailable raw data. Second, the model assumes ideal circuit behavior, omitting real-world imperfections like noise or op-amp saturation, which impact chaos. Third, gradient estimation via finite differences is sensitive to hyperparameters, risking convergence to local minima. Lastly, calibration is computationally intensive, with simulation count scaling with parameters and epochs, limiting real-time use and scalability for complex models or extended data.

\section{Future Work}
Future research directions include:
\begin{itemize}
    \item Tune additional physical parameters such as resistance \( R \), capacitance \( C \), and the shape of the nonlinear diode function \( u(x) \) to improve physical realism.
    \item Integrate direct hardware interfacing to enable real-time data acquisition and feedback, reducing reliance on digitized samples.
    \item Improve the model by refining $u(x)$ to better reflect real diode behavior and incorporating experimentally validated non-idealities.
    \item Use Lyapunov exponents and other chaos quantifiers as metrics to assess model accuracy and dynamical fidelity.
    \item Replace finite-difference gradient estimation with adjoint sensitivity analysis or automatic differentiation for efficiency.
    \item Leverage parallelism or GPU acceleration for faster parameter sweeps.
\end{itemize}

\section{Supplementary Material}
The code is provided at \url{https://github.com/Akadib/SGD_Calibration_Sprott_Circuit}. Extracted data will be available upon request. All other methods are described within the main text.

\section{Acknowledgments}
We express our sincere gratitude to several individuals who significantly contributed to this project. Dr. Todd Neller suggested the application of Stochastic Gradient Descent (SGD) for the calibration process, which greatly enhanced our methodology. Finally, Dr. Bret Crawford offered valuable feedback on nonlinear circuit modeling, enriching our understanding of the system’s dynamics.

\section{Author Declarations}
\subsection{Conflict of Interest}
The authors have no conflicts to disclose.

\subsection{Ethics Approval}
This study did not involve experiments using human or animal subjects, so no ethics approval is required.

\subsection{Author Contributions}

\textbf{Adib Kabir}: Conceptualization (equal), Data curation (equal), Formal analysis (equal), Investigation (equal), Methodology (equal), Software (equal), Writing – original draft (equal). \\\textbf{Onil Morshed}: Data curation (equal), Formal analysis (equal), Investigation (equal), Methodology (equal), Software (equal), Writing – original draft (equal). \\\textbf{Oishi Kabir}: Conceptualization (equal), Data curation (equal), Formal analysis (equal), Investigation (equal), Methodology (equal), Software (equal), Writing – original draft (equal).
\\\textbf{Juthi Hira}: Conceptualization (equal), Data curation (equal), Formal analysis (equal), Investigation (equal), Methodology (equal), Software (equal), Writing – original draft (equal).
\\\textbf{Dr. Caitlin Hult}: Supervision (supportive), Mathematical modeling (review), Writing – review and editing (minor). \\

\section{Data Availability Statement}
The data that support the findings of this study are available from the corresponding author upon reasonable request.

\appendices
\section{Computational Expense Considerations}\label{app:comp_cost}
Optimizing parameters in chaotic systems—especially using orders like orders—requires numerous simulations due to sensitivity to changes. Our finite-difference simulation requires two evaluations per parameter per iteration. Let:
\begin{itemize}
    \item \( d \): Number of parameters (e.g., \( T_{\text{con}}, K, R_v \)),
    \item \( E \): Number of SGD iterations,
    \item \( T \): Time steps in each simulation,
    \item \( \Delta u \): Integration step size.
\end{itemize}

Each gradient requires two simulations per parameter, so total simulations are:
\begin{equation}
N_{\text{sim}} = 2d \cdot E.
\end{equation}
With complexity:
\[
\mathcal{O}(2dET) = \mathcal{O}(dET).
\]
For \( d=3 \), \( E=100 \), \( T=284 \):
\[
N_{\text{sim}} = 600, \quad \mathcal{O}(1.7 \times 10^5).
\]
Adding two parameters (\( d=5 \)):
\[
N_{\text{sim}} = 1000, \quad \mathcal{O}(2.84 \times 10^5),
\]
a 67\% increase. Expanding the cost function with more terms raises costs linearly, necessitating careful justification for added complexity.

\end{document}